\documentclass[aps,pre,showpacs,twocolumn]{revtex4}

\usepackage{graphicx}
\usepackage{dcolumn}
\usepackage{bm}

\begin{document}
\newcommand{\gam}{\dot{\gamma}}
\newcommand{\eps}{\epsilon^*}

\preprint{APS/123-QED}

\title{Comparison of Molecular Dynamics with Hybrid Continuum-Molecular Dynamics
for a Single Tethered Polymer in a Solvent}

\author{Sandra Barsky}
\email{s.barsky@ucl.ac.uk}
\author{Rafael Delgado-Buscalioni}
\email{uccarde@ucl.ac.uk}
\author{Peter V. Coveney}
 \email{P.V.Coveney@ucl.ac.uk}
\affiliation{Centre for Computational Science, Dept of Chemistry,
University College London, London UK, WC1H 0AJ
}%

\date{\today}

\begin{abstract}
We compare a newly developed hybrid simulation method which combines
classical molecular dynamics (MD) and computational fluid dynamics
(CFD) to a simulation consisting only of molecular dynamics. The
hybrid code is composed of three regions: a classical MD region, a
continuum domain where the dynamical equations are solved by standard CFD
methods, and an overlap domain where transport information from the other two
domains is exchanged. The exchange of information in the overlap
region ensures that momentum, energy and mass are  conserved.
The validity of the hybrid code is demonstrated by studying a single
polymer tethered to a hard wall immersed in explicit solvent and undergoing
shear flow. In classical molecular dynamics simulation a great deal
of computational time is devoted to simulating solvent molecules,
although the solvent itself is of no direct interest. By contrast,
the hybrid code simulates the polymer and surrounding solvent
explicitly, whereas the solvent farther away from the polymer is
modeled using a continuum description. 
In the hybrid simulations the MD domain is an open system 
whose number of particles is controlled to filter the perturbative density
waves produced by the polymer motion. 
We compare conformational
properties of the polymer in both simulations for various shear
rates. In all cases polymer properties compare extremely well between
the two simulation scenarios, thereby demonstrating that this hybrid
method is a useful way to model a system with polymers and under
nonzero flow conditions. There is also good agreement between the MD and
hybrid schemes and experimental data on tethered DNA in flow.
The computational cost of the hybrid protocol can be reduced to less than
$6\%$ of the cost of updating the MD forces, confirming the
practical value of the method.

\end{abstract}

\maketitle

\section{\label{sec:intro}Introduction }

Molecular dynamics (MD) simulations have long been used to model
complex fluids both in and out of equilibrium.  As computers get more
powerful there has been an increasing desire for more chemically
accurate models of these fluids. This means that simulations are
becoming larger and more accurate, but also that much simulation time
is being devoted to model in detail parts of the computational system
of little direct scientific interest. Hybrid methods combine regions of relatively high
degree of chemical accuracy in a specific region of interest, and a
more coarse-grained model where the dynamics can be solved in a less
computationally intensive way, further away from the specific region
of interest. We focus, in particular, on hybrid methods that combine
Lennard-Jones type specificity with larger scale continuum
methods. Such hybrid methods have been applied in a number of fields,
including Lennard-Jones fluids \cite{thomp}, biophysics \cite{ibm} and
MD/CFD coupling \cite{li,li2}. This type of simulation technique is
particularly useful in studying interface problems, where the region
of interest is a localized part of the entire system.

Typical hybrid methods consist of three regions: a traditional region
where dynamics are simulated using well-established techniques such as
molecular dynamics \cite{allen}, a continuum region where CFD or
elasticity differential equations are solved using classical
techniques, and an overlap region where the necessary transport
information of the MD and continuum regions are exchanged.  The
primary motivation for using a hybrid scheme is to reduce computer
time devoted to simulating bulk regions of little direct interest. As
such, a hybrid scheme is ideally suited to studying interfacial
systems.

In this paper we apply a hybrid technique to a single polymer tethered
to a wall with explicit solvent.  The complex dynamics arising from
this system have attracted a degree of interest from experimentalists,
who used fluorescence microscopy and
videomicroscopy to investigate the dynamic properties of individual
DNA chains in a shear flow, either tethered to a wall \cite{doyle} 
or free \cite{leduc}. These experiments
reveal that the structural quantities, such as the mean elongation of the
polymer, are very sensitive to flow environment and that the dynamical
properties depend strongly on the initial conformation. Moreover, 
care needs to be taken to control the finite size effects, such 
as those due to 
long-ranged hydrodynamic interaction between the polymer and the walls \cite{leduc}. This large ``sensitivity'' of the tethered polymer
dynamics is in fact a valuable test for the hybrid model. First, the
hybrid model reduces the size of the MD simulation box while 
avoiding finite size effects and, second, the coupling has to be able
to perfectly reproduce flows at very small shear rates.  As shown in
recent work \cite{BusF}, this second task is non-trivial because the
signal-to-noise ratio of the stress that one needs to communicate from
the particle to the continuum system is very small.

The problem of tethered polymers under flow has a geometry which is
ideal for a hybrid scheme. The scientific interest lies around
the polymer although, in a standard MD simulation, the solvent
particles within the bulk flow require most of the computational
time. Single polymers in a bath of explicit molecular solvent have
been the focus of a great deal of attention in the last decade
\cite{dunw2,dunw,aust,pr95,dunw3,dep}.  Many of these studies are
devoted to examining a free chain in solution in order to make
comparisons with theoretical predictions, explore the dynamics regime
beyond the short-time dynamics or extract scaling laws as a function
of polymer length or shear flow.  In these studies, the solvent is
explicity simulated.  For example, the study by D\"{u}nweg and Kremer
\cite{dunw} uses a polymer of length $L=60$ beads in a bath of $7940$
Lennard-Jones spheres.  Aust, Kr\"{o}ger and Hess \cite{aust} simulate
polymers of length $L=10$ to $60$ in systems where the total number of
particles including solvent ranges from $1000$ to $5832$.  It is clear
in these cases that most of the computational effort is devoted to
solving the equations of motion of the solvent particles when the real
scientific interest lies in the polymer behavior.

The single polymer we study is tethered to a wall, and  a variety of shear
rates is imposed as a model interfacial problem to compare classical MD
techniques to the hybrid simulation.
In classical MD, we sandwich the polymer and
solution between two explicit walls, and impose periodic boundary
conditions in the remaining two directions. The polymer is tethered to
the bottom wall, and shear is created by moving the top wall at
constant velocity in a direction parallel to the wall. In the hybrid
case, we model one wall, the polymer and some of the solvent
explicitly using MD, and impose shear by a boundary condition in the
CFD regime of the calculation. The shear is translated to the MD
regime via energy and momentum flux transfers in the overlap region.
We compare various conformational properties of the polymer for the
two techniques.

Our paper is organized as follows. In the following section we
briefly outline both the classical MD simulation  and the hybrid
simulation techniques. In Sec. \ref{sec:result} we compare the
conformation of the polymer as calculated by each simulation method. 
The computational costs and benefits of the hybrid scheme are compared to
 classical MD. We also compare our results to experimental data of tethered
DNA under shear flow. We
conclude with a discussion in Sec. \ref{sec:con}.

\section{\label{sec:method} Method}
We describe in this section both the molecular dynamics and hybrid
dynamics models used in our simulations. The MD part of the hybrid
scheme was the same as the classical MD used in the pure molecular
dynamics simulations.

\subsubsection{Molecular Dynamics}
The polymer model and simulation techniques are similar to those used
in previous work \cite{robbins,barsky}. The polymer potential is based
on the bead-spring model developed by Kremer and Grest \cite{grest1}.
Linear polymers containing $N=60$ beads each are created
by linking nearest  neighbors on a chain with
the potential \begin{equation}\label{vgrest}
U_{nn}(r_{ij})=\left\{\begin{array}{ll}-\frac{1}{2}kR_0^2\ln\left[1-
\left(r_{ij}/R_0\right)^2\right]&r_{ij}<R_0\\ \infty & r_{ij}\geq
R_0\,, \end{array}\right.\end{equation} where $r_{ij}$ is the distance
between beads $i$ and $j$, $R_0=1.5\sigma$, $k=30\epsilon/\sigma^2$,
and $\sigma$ and $\epsilon$ set the length and energy scales,
respectively. The monomers in the solvent and in the polymer  interact
through a truncated Lennard-Jones (LJ) potential \begin{equation}\label{LJ}
U_{LJ}(r_{ij})=\left\{\begin{array}{ll}4\epsilon\left[\left(
\sigma/r_{ij} \right)^{12}-\left(\sigma/r_{ij}\right)^{6}
\right]&r_{ij}<r_c\\ 0&r_{ij}\geq r_c\,. \end{array} \right.
\end{equation}
The cutoff is set at $r_{c}=2^{1/6} \sigma$ for all fluid particles to
produce a purely repulsive interaction between beads.

The bounds of the simulation cell are periodic in the $x$ and $y$
directions, with periods $L_x \cong 38.5 \sigma$ and $L_y \cong 33.4
\sigma$, respectively. In the $z$ direction the cell is bounded by top
and bottom walls. Each wall contains two layers of $1600$ spheres
strongly tethered  to the sites of a $(1,1,1)$ plane of an fcc lattice
by harmonic springs of stiffness $\kappa=1320 \epsilon \sigma^{-2}$.
The wall atoms do not interact with each other, and the wall-fluid
interaction is LJ with an increased cutoff of $r_c=1.25 \sigma$ and
increased energy scale of $\epsilon_{wf} = \sqrt{1.7} \epsilon$. The
increased cutoff and energy ensure sufficient adhesion of the fluid to
the wall so that the slip at the wall is minimized for the shear rates
considered here. The polymer is anchored to the wall by enforcing the
tethering potential, Eq. (\ref{vgrest}) between the end of the
polymer and one wall atom.

The walls are  $48 \sigma$ apart for the pure MD simulation; in the hybrid
simulation the molecular dynamics region persists for $19
\sigma$.
There are sufficient solvent
monomers to yield a mean fluid density of approximately 
$\rho= 0.8\sigma^{-3}$ in the
center of the simulation cell, although density oscillations are
induced within a few $\sigma$ of the walls \cite{khare}.

The equations of motion are integrated using a velocity Verlet
algorithm \cite{allen}, with a time step 
$\delta t = 0.0075 \tau$, where
$\tau = \sigma \sqrt{m/\epsilon}$ is the basic unit of time, and $m$ is
the mass of a monomer. A constant temperature of $k_BT=1.0\epsilon$ is
maintained with a Langevin thermostat \cite{grest1}. To ensure that
this thermostat does not bias the shear profile, the Gaussian white
noise and damping terms are only added to the equations of motion for
the velocity components normal to the mean flow, that is the $y$ and $z$
directions  \cite{robbins}.

The shear flow in our pure MD simulation  is induced by moving
the atomistic top wall at a constant speed $v_{x}$ in the $x$ direction.  In the
hybrid simulation, a shear boundary condition is used for the continuum
regime, and this resulted in shear in the MD regime by the exchange of
momentum in the overlap domain. The starting configuration was that
of a single polymer tethered to the wall, in an equilibriated solvent.
We repeat the simulation for two different starting configurations,
{\it i.e.} each configuration has a polymer tethered to a different
wall atom, and the initial conformation of the polymer is different.
The initial polymer configurations were either taken from previous
simulations on melts \cite{barsky}, or generated from a random walk.
 Although over long periods of time we
expect that different starting configurations will give the same configurational averages,
 previous work \cite{dunw} has shown that hundreds of
different initial configurations are required to arrive at reasonable ensemble
averages. In view of this, we used two initial configurations for the pure MD simulations;
although this falls short of the number of initial configurations required to achieve
ensemble averages, it does give us a window over which to compare the hybrid simulation.

The local shear rate, $\dot{\gamma}$, of the fluid is  calculated by
computing the local change in the $x$ component of velocity, $v_x$, as a
function of $z$, {\it i.e.} $\dot{\gamma}= \partial v_x/ \partial z$.
The upper wall velocity was chosen so that the shear rate
$\dot{\gamma}$ assumed the values $ 0.0, 0.0005, 0.001, 0.002, 0.005, 0.01 \tau^{-1}$. Simulations
at higher shear rates were created from lower shear rates by increasing
the wall velocity or boundary condition and allowing the system to
achieve steady state. The simulations were done for at least one
million time steps, and the runs of  $\dot{\gamma}=  0.001  \tau^{-1}$ and
$0.005 \tau^{-1}$ were simulated for at least ten million time steps,
corresponding to a total run time of $75,000 \tau$. In the analysis in
the following section the first $250,000$ time steps of data for each
given shear rate were ignored, to allow the system to reach steady
state; this length of time was determined to be the longest time
necessary for the system to reach steady state once a new shear rate was imposed.

\subsubsection{Hybrid Dynamics}
\label{sec:hybrid}  

Our hybrid dynamics code \cite{usher} consists of three domains: the particle
domain (P) which was by the same molecular dynamics method described
above, the continuum domain (C) treated by standard continuum fluid
dynamics and, an overlap region where information from the other two
domains is exchanged.
The hybrid scheme is a protocol to exchange fluxes of conserved quantities,
specifically mass, momentum and energy between both classically treated regimes.
To implement the two-way flux exchange, the overlap region consists of two different subdomains:
the P$\rightarrow$C and the C$\rightarrow$P cells.
Within the C$\rightarrow$P region, the fluxes from the continuum
domain are imposed on the particle domain, whereas within
the P$\rightarrow$C cell the microscopic fluxes are coarse-grained in time and space \cite{usher}
to supply boundary conditions for the continuum domain.

\begin{figure}[hbt]
\includegraphics[width=10cm,totalheight=10cm]{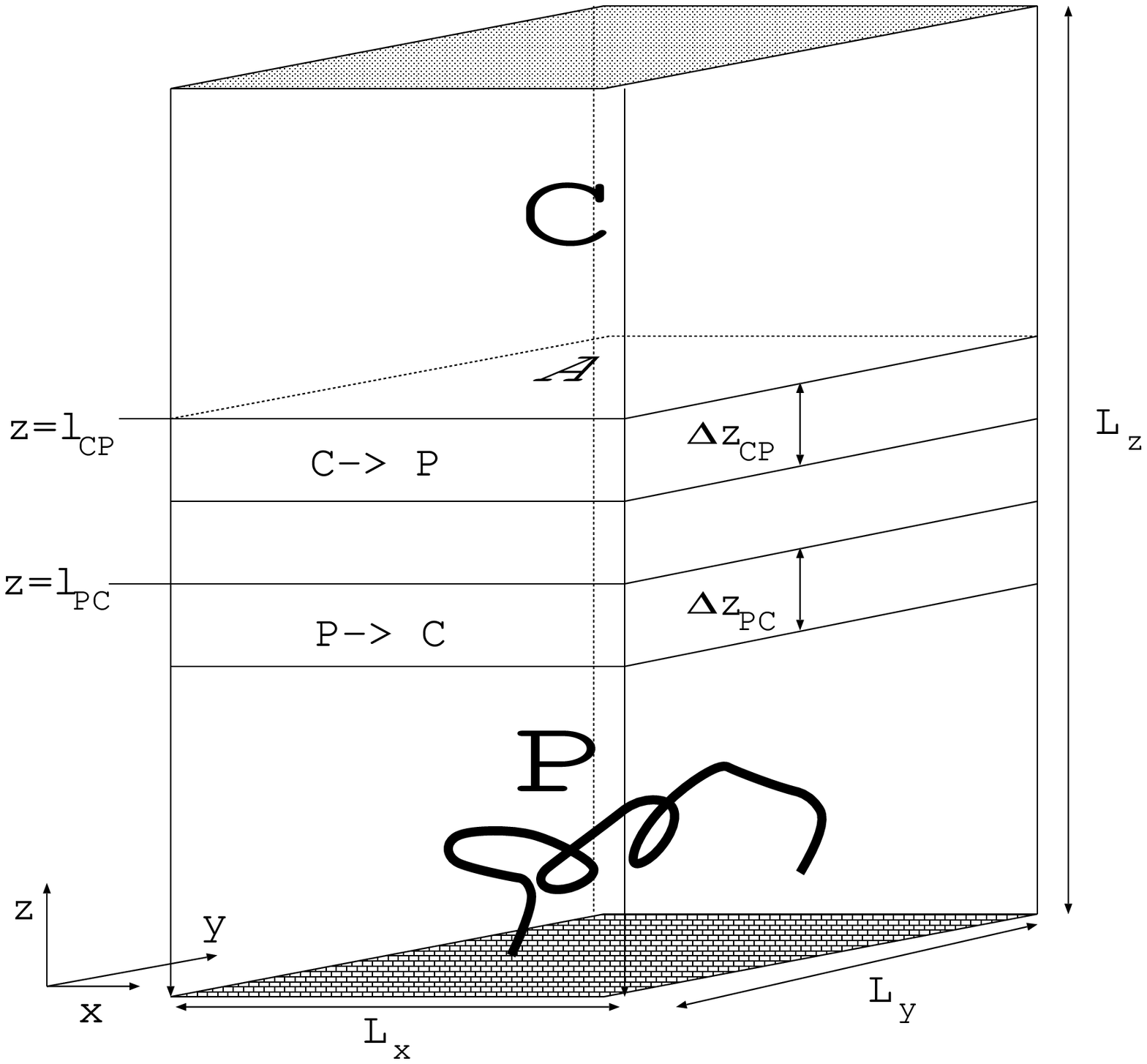}
\caption{The domain decomposition of the hybrid scheme. The polymer is
embedded within the particle region (P) which is described by molecular dynamics,
including the atomistic (lower) wall and the solvent (Lennard-Jones
particles). Fluid flow within the continuum region (C) is described by
an unsteady Stokes equation and is solved using finite volumes. The
handshaking region contains the C$\rightarrow$P  and the
P$\rightarrow$C cell, where the two-way exchange of information
is established. The P and C domains overlap within $l_{PC}\leq z\leq
l_{CP}$. The area of the P$\rightarrow$C and C$\rightarrow$P cells is
the surface of the system in the periodic directions, $A=L_xL_y$.
The Couette flow moves along the $x$ direction driven by the velocity
imposed by the upper boundary condition, which corresponds to the upper wall
velocity in the pure MD simulations, $u_{wall}$. The magnitudes of each
length shown in the figure are given in Sec. \ref{sec:hybrid}. }
\label{overlap}
\end{figure}

The spatial decomposition used for the present set-up is shown in
Fig. \ref{overlap}. The molecular dynamics domain
ranges from the atomistic wall at $z\simeq 0$
and extends to $z=l_{CP}=19\sigma$. The continuum fluid dynamics domain comprises
$l_{PC}\leq z \leq L_z$, where $l_{PC}\simeq 14.5\sigma$ is the
$z-$coordinate of the P$\rightarrow$C interface and $L_z=50\sigma$
is the extent of the whole simulation domain.
The center of the P$\rightarrow$C cell is located at $z=13.4\sigma$;
it has a volume $V_{PC} =\Delta z_{PC} A$ where $A=L_x\times L_y$
and $\Delta z_{PC}\simeq 2.2 \sigma$ is the extension along the $z$ direction.
The C$\rightarrow$P cell is placed at a distance $2.2\sigma$  from the end of
the P$\rightarrow$C cell and covers a region of
 $\Delta z_{CP}\simeq 2.2 \sigma$, from $z\simeq 16.8\sigma$
to $z=l_{CP} \simeq 19\sigma$.  

In what follows we outline the coupling protocol and provide the
numerical details used in the present implementation.  The
C$\rightarrow$P coupling represents the most complicated part of the
hybrid scheme; a more
detailed explanation of the method in the frame of the general case of
unsteady flows with mass, momentum and energy exchanges can be found
 in reference \cite{BusH1}. The steady flow
considered here only carries momentum  along the $x$
direction. Although the mean flux of mass and energy across the C and P
interfaces is zero, fluctuations in the particle system
produce perturbative mass currents along the $z$ direction which need to be
taken into account.  This part of the C$\rightarrow$P scheme is presented in Appendix \ref{sec:appx}.

We now focus on how the momentum flux is exchanged between the C and P
domains, starting with a discussion of the C$\rightarrow$P coupling.
For the pure Couette shear flow considered here,
the momentum flux due to the C flow across any $z=$constant surface
is given by
\begin{equation}{\bf \Pi}=P {\bf k} -\eta \gam {\bf i}, \end{equation}
 where $P=P(\rho,T)$ is the
hydrostatic pressure, $\eta$ is the dynamic viscosity and $\gam\equiv
\partial v_x/\partial z$ is the shear rate.
The value of the dynamic viscosity
was measured in a previous pure MD simulation via
the standard non-equilibrium procedure \cite{aust,barsky,Evans};
for $\rho=0.8$ and $T=1.0$ we obtained  $\eta=1.75 \pm 0.04$.
 The stress induced by the C-flow in the P domain is given by the
local momentum flux at the C$\rightarrow$P interface, ${\bf
\Pi_{CP}}$. In order to introduce this stress into the molecular
dynamics domain we add an  external force ${\bf F_{ext}}=-{\bf
\Pi_{CP}} A$ to those molecules within the C$\rightarrow$P cell.  At
any instant of time, $t$, this force is equally distributed among the
$N_{CP}(t)$ particles inside the C$\rightarrow$P cell, so the external
force per particle is ${\bf F_{ext}}/N_{PC}=-{\bf
\Pi_{CP}}\,A/N_{CP}$.  Note that this external force has a component
normal to the C$\rightarrow$P interface, which provides the
hydrostatic pressure, and a  tangential component providing the shear
stress.  The molecules are free to enter or leave the C$\rightarrow$P
region, so the number of molecules within this region, $N_{CP}(t)$, and
the value of the overall external force fluctuate in time. The average
``pressure force'' per particle is $P(\rho,T)/(A \bar{N}_{CP})$, where
$\bar{N}_{CP}\sim 2000$ is the mean number of particles within the
C$\rightarrow$P cell (see Appendix \ref{sec:appx}), $A= L_x \times L_y =
1286\sigma^2$ and the pressure $P(\rho,T)$ is given from the equation
of state provided by Hess {\em et al.}  \cite{eosWCA}: $P(0.8,1)\simeq
6.5\epsilon/\sigma^3$.  Such a force prevents the escape of
particles and, although it induces some ripples on the density profile
over the C$\rightarrow$P cell, it maintains the correct value of the
density along the inner part of the MD domain, as seen in
Fig. \ref{density}b and discussed further in the Appendix \ref{sec:appx}.

The shear force is distributed over the particles
in the same way as described above for the pressure force.
In this case, the flux of $x$-momentum to be injected in the particle
system is $\eta \gam_{CP}$, where $\gam_{CP}$ is the local shear rate
of the C-flow measured at the C$\rightarrow$P interface.  

We next discuss the P$\rightarrow$C coupling.  The continuum domain is
a coarse-grained description of the fluid, therefore any information
transferred from the molecular to the continuum system needs to be
averaged in space and time.
These averages need to be local in the continuum space and time
coordinates.  To that end, the particle quantities are averaged within
the P$\rightarrow$C cell and over a time interval $\Delta t_{av}$.  It
is important to stress that within the P$\rightarrow$C cell each
particle's dynamics are not directly modified by any external
artifact, in other words the motion of each particle is uniquely
determined by the usual molecular dynamics scheme.
To ensure consistency within the hybrid scheme, $\Delta t_{av}$ and the volume
of the P$\rightarrow$C cell are restricted 
\footnote{As explained in
references \cite{BusH1,BusPHIL}, these restrictions come from the C flow
(stability of the scheme and spatio-temporal resolution) and from the P system
(local thermodynamic equilibrium, signal-to-noise ratio).}.
 For the  steady flow employed in this study,
the most compelling condition is to guarantee that the signal-to-noise
ratio of the momentum flux is larger than one and for that  reason
$\Delta t_{av}$ needs to increase as $\gam$ decreases
\cite{BusF}. We used $\Delta t_{av}=100 \tau$ for $\gam \leq 0.001$
and reduced it gradually to $10\tau$ for the fastest flows considered.

To solve the equations of motion in the continuum domain we  used
the finite volume formulation \cite{Patankar} because it matches by construction
the fluxes across cells.  Since the solvent flow is isothermal,
incompressible and there is  a uniform pressure,
the mean $x$-velocity is governed by $\partial
v_x/\partial t=\nu \partial^2 v_x/\partial z^2$, where $\nu=\eta/\rho$ is the
kinematic viscosity and $v_x$ is the velocity in the $x$ direction.
At the top of the simulation cell we impose a smooth wall in the CFD
sense.  This wall moves at a constant
velocity $v_x(L_z,t)=u_{wall}$ which creates the shear flow in the
simulation.  The protocol for the P$\rightarrow$C coupling establishes
the boundary condition for the continuum domain at the P$\rightarrow$C
interface, $z=l_{PC}$.  The coarse-grained microscopic flux of
$x$-momentum across the P$\rightarrow$C interface, whose expression is
given in Ref. \cite{BusH1}, is set equal to the corresponding value
for the C flow at the $z=l_{PC}$ boundary, $\eta \gam_{PC}$, where
$\gam_{PC} \simeq (v_x(l_{PC}+\Delta z)-v_x(l_{PC}))/\Delta z$.  This
condition gives the desired velocity to be imposed at the boundary
$v_x(l_{PC})$.  The continuity of velocity is ensured by adding a
relaxing term in the flux equation which drives the C-velocity at the
interface towards the corresponding averaged P-velocity (see
references \cite{BusPHIL,BusF} for details).


\section{\label{sec:result} Results and Discussion}

In this section we compare the conformational behavior of the
polymers from  the MD and hybrid simulations; we use two independent MD
simulations for comparison. We study two MD systems because it is well known
\cite{dep} that two simulations, or experiments, on a tethered polymer
may exhibit considerable variation in conformational behavior, even at rather high shear rates.
We conclude this section with a discussion of the computational costs
and benefits of the hybrid and classical MD techniques.

\begin{figure}[hbt]
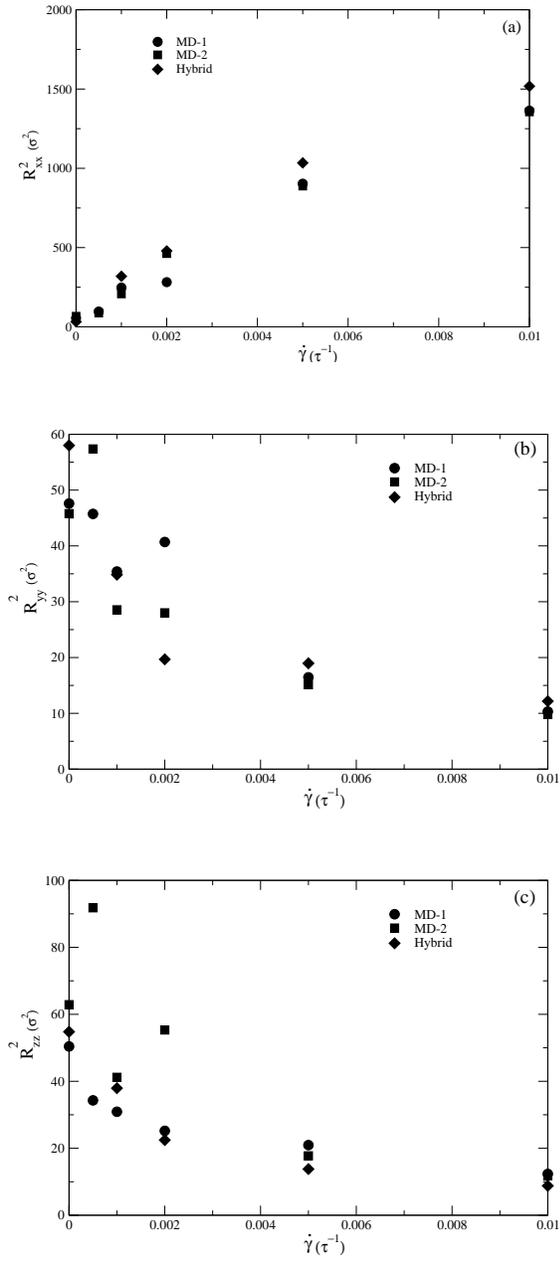

\includegraphics[totalheight=5.2cm ]{rxx.eps}
\vspace{0.7cm} \\
\includegraphics[totalheight=5.2cm ]{ryy.eps}
\vspace{0.7cm} \\
\includegraphics[totalheight=5.2cm ]{rzz.eps}
\caption{The  $x$, $y$ and $z$  components of the mean square
end-to-end vector, $R^2$, are shown as a function of shear rate for two
independent MD simulations and one hybrid. Error bars, not shown, are
approximately $15 \%$. The $x$-component of $R^2$ increases as the shear rate
increases, while the $y$ and $z$ components decrease. At low shear rate
the hybrid simulation is
well within the variation of the MD simulations. At the highest shear rate
the values for $R^2_{xx}$ agree within the measured uncertainty.}
\label{end-x} 
\end{figure}

In Figure \ref{end-x} we show the mean-square end-to-end distance
$R^2$ of the polymer in each of the $x,y$ and $z$ directions. This is
a standard measure of polymer conformation \cite{doi}, and its values
are related to the values of the radius of gyration.  At low shear
rates the polymer conformation calculated from the hybrid simulation
is well within the measured conformations of the MD simulations. At
the highest shear rates, Fig. \ref{end-x}(a) shows the conformation of
the hybrid polymer to be about $10\%$ larger than the polymer in the
MD simulation; this difference is within the standard deviation of
$R^2_{xx}$, which is about $15\%$. Figs.\ref{end-x}(b) and (c) show
the $y$ and $z$ components of $R^2$ as a function of shear rate
$\dot{\gamma}$. In both cases the two MD simulations serve as good
indicators of the variability of the conformational behavior of a
single polymer; the conformation of the polymer in the hybrid
simulation is well within the variations found in the two MD
simulations, at all shear rates.

\begin{figure}[hbt]
\includegraphics[width=5cm,totalheight=10cm]{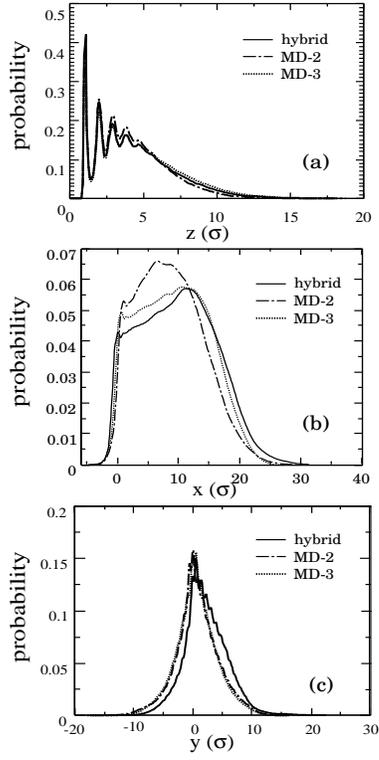}
\caption{Probability of finding a monomer in the $x$ (a), $y$ (b) and
$z$ (c) coordinates in a flow with shear rate $\gam=0.001 \tau^{-1}$.
Comparison is made between the result obtained with the hybrid scheme
and the outcome of two pure MD simulations with different initial
conditions.  
}
\label{histo1d}
\end{figure}

Figure \ref{histo1d} shows the probability of the maximum extension from
as a function of distance along $x, y$ and $z$ directions, for a shear rate of
$\dot{\gamma}=0.001 \tau^{-1}$. It is clear that the variation of
the distributions obtained  with the hybrid simulation is
well within the distribution of the two MD simulations. This indicates
that not only is the average conformation comparable between the two
simulation techniques, but that the probability distributions also
compare favorably.

\begin{figure}[hbt] 
\includegraphics[width=10cm,totalheight=10cm]{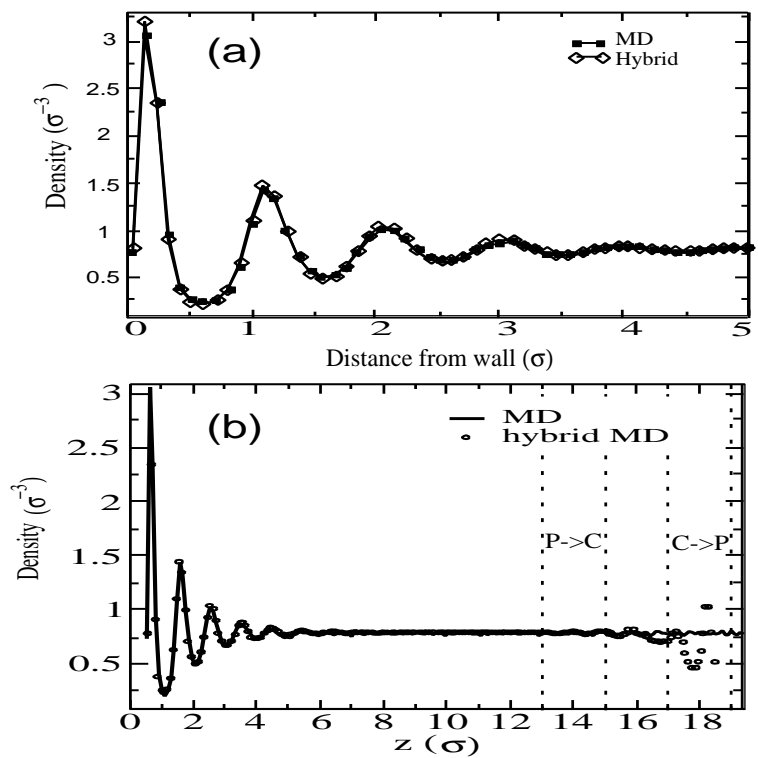}
\caption{Monomer and solvent density as a function of distance (z) from
the wall, for shear rate $\dot{\gamma}=0.005 \tau^{-1}$. (a) shows the 
density fluctuations near the lower wall. 
In (b) dashed lines indicate the locations of
the coupling buffers used in the hybrid scheme, 
P$\rightarrow$C and C$\rightarrow$P.
Outside the $C\rightarrow$P region, the monomer
density is unaffected by the hybrid scheme}
\label{density}
\end{figure}

In Figure \ref{density} we show the density, of both the solution
monomers and polymer, $\rho= N/V$ as a function of distance from the
wall for both the MD and hybrid simulations. The density is calculated
in slices of approximately $0.01 \sigma$ perpendicular to the
wall. The regular spacing of the wall monomers, as two monolayers of a
$(1,1,1)$ face of an fcc crystal, induces an ordering in the fluid;
this ordering is well established \cite{robbins} and persists for
approximately $5 \sigma$. At the wall the monomer density variations
are identical for both the MD and hybrid simulations as seen in Fig
\ref{density}(a). In Fig. \ref{density}(b) we see that the density of both
simulations remains the same until the monomers in the hybrid system
feel the effects of the constant pressure condition imposed on the
overlap region. The constant pressure is implemented as a simple
normal force per particle on all monomers in the C$\rightarrow$P
regime, as discussed in the previous section. This force induces a
local ordering in the monomers, which in turn creates density
fluctuations. It is noteworthy, however, that these density
oscillations are much lower than at the atomistic wall, shown for
comparison.  More recent work on the hybrid scheme has established
that we can reduce these density fluctations even further, as
discussed in the Appendix \ref{sec:appx}.  In the MD simulation, the
upper wall is identical to the lower one, and thus the density
fluctuations near the former are the same as those at the latter.

\begin{figure}[hbt]
\includegraphics[width=10cm,totalheight=10cm]{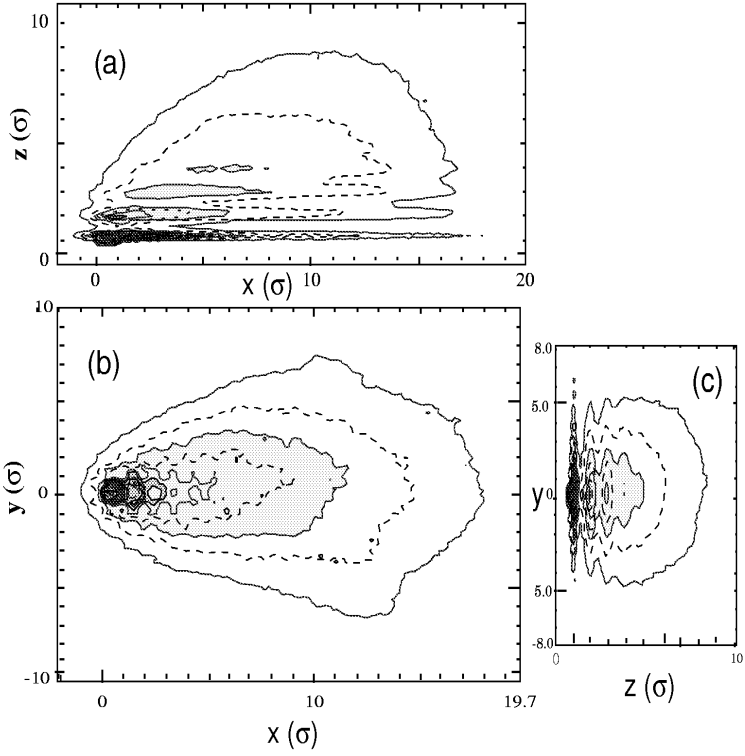}
\caption{Probability of finding a monomer
in the $x-z$ (a), $x-y$ (b) and $y-z$ (c) planes in a flow with shear rate
$\gam = 0.001 \tau^{-1}$. The maximum of the probability
distribution is located near the attachment site. The shaded region
corresponds to an iso-probability value of $0.021$ and the values
of consecutive iso-probability contour lines are separated by $0.01$.
The histograms were obtained from the calculation of
a pure MD simulation with a total simulation time of 
$78750 \tau$.}
\label{histo2d}
\end{figure}

Fig. \ref{histo2d} shows the probability of finding any polymer bead
in a plane, where the  plane slices are $0.2 \sigma$ in thickness. The
two-dimensional probabilities were calculated in an analogous way to
the one-dimensional probabilities discussed above. The shear rate shown
is $\gam=0.001$.  Inspection of the two-dimensional bead distributions
indicates that below a distance of $\sim 5\sigma$ to the wall, the
beads tend to be ordered in layers parallel to the wall plane.  This
result is not only a consequence of the polymer-wall interaction but
also an effect of the interaction with the solvent.  Near the wall the
solvent is ordered in layers, as in Fig. \ref{density}(a), and the
polymer minimizes the monomer-solvent potential energy by adapting its
distribution to match the locations of the solvent layers.  The order
induced by the wall in the polymer structure can be noticed even in
the isovalues of the probability distribution along the wall plane
$x-y$, shown in Fig. \ref{histo2d}(b), and along the $z-y$ plane in
Fig. \ref{histo2d}(c).  Over a distance of $\sim 6\sigma$ around the
attachment position the isovalues of the probability distribution in
the $x-y$ plane delineate the minimum energy lines of the wall atoms
LJ potential.  In this model, the size of the wall atoms was chosen to
be the same as those of the monomers and solvent particles $1\sigma$.
In view of Fig. \ref{histo2d}(b), one should expect that the structure
of the polymer is quite sensitive to any modification in the details
of the wall-fluid interaction, owing to either changes in the size of
the wall atoms or details of the interaction potential.

\begin{figure}[hbt]
\includegraphics[width=8cm,totalheight=10cm,angle=-90]{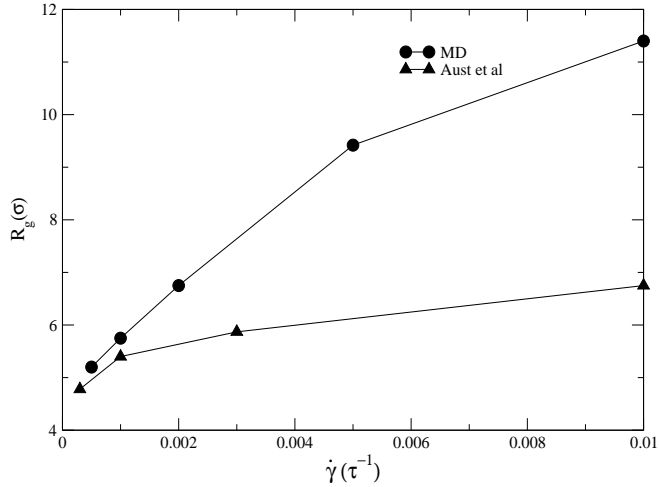}
\caption{Comparison of the radius of gyration of the polymer as calculated by MD simulation
 in this paper, and that of Aust {\it et al.} \cite{aust}.
The higher $R_g$ found in this
work is due to the attraction between the wall and the polymer.}
\label{rg}
\end{figure}

In Fig. \ref{rg} we present a comparison of the radius of gyration
$R_g$, as calculated from MD simulations in this work and that of
Aust, Kr\"{o}ger and Hess \cite{aust} (AKR) who studied a single free
polymer in a bath of solvent molecules, at a variety of imposed shear
rates.  The potential used to describe the polymer and solvent were
the same in both AKR's work and ours; however AKR used a slightly
higher density, $\rho=0.85 \sigma^{-3}$, compared to our value of $\rho=0.8
\sigma^{-3}$. The simulation of AKR used no walls, so the polymer was
free to respond to the imposed shear so as to best lower the free energy
of the system. Hence the usefulness of the comparison lies primarily
in exploring the effect of the wall on the polymer.  We see that at
extremely low shear rates the values of the radii of gyration are
quite comparable. As the shear rate increases the value of $R_g$ that
we calculate becomes much larger than for the equivalent free
polymer. This is due entirely to the attraction the polymer has with
the wall. 

\begin{figure}[hbt]
\includegraphics[width=8cm,totalheight=12cm,angle=-90]{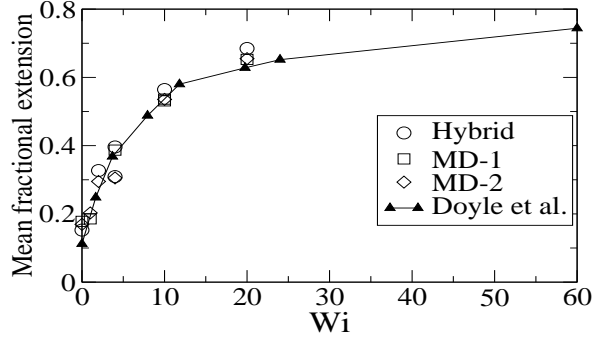}
\caption{The fractional elongation along the flow direction
versus the Weissemberg number, Wi=$\tau_0\gamma$.
The longest decay time at zero shear rate 
obtained from our data is $\tau_0=2000\tau$.
Comparison is made with the experimental results of 
Doyle {\em et al.} \cite{doyle} on tethered polymers.}
\label{elongwi}
\end{figure}

Our results are in agreement with the experimental  
findings of Doyle {\em et al.} \cite{doyle} for individual tethered  
DNA chains under shear flow.  
For a quantitative comparison with these experimental data we   
evaluated  the Weissemberg number, Wi, defined as the product of  
the shear rate and the longest relaxation time of the polymer,  
{\em i.e.} the relaxation time at zero shear rate $\tau_0$.  
We calculate $\tau_0$ from the autocorrelation of the polymer extension at  
$\gam=0$ and obtain $\tau_0 \simeq 2000 \tau$.
Also, the fractional extension is calculated by  
normalizing the polymer extension with its contour length: $0.965\times  
(N-1)$, where $N=60$ is the the number of monomers and  
$0.965\sigma$ is the mean separation between two consecutive  
beads \footnote{We checked that the monomer separation remains  
unchanged for the range of $\gam$ considered (it varies in less than  
$0.005\sigma$.)}. Using this value of  
$\tau_0$ we plot in Fig. \ref{elongwi} the normalized mean fractional extension along  
the flow direction versus the Weissemberg number, along with the   
the expermimental results of Doyle {\em et al}. The results obtained with
both, the MD and hybrid simulations 
are in very good agreement with the experimental data
for the range of shear rates considered here.

\begin{figure}[hbt]
\includegraphics[width=8cm,totalheight=12cm,angle=-90]{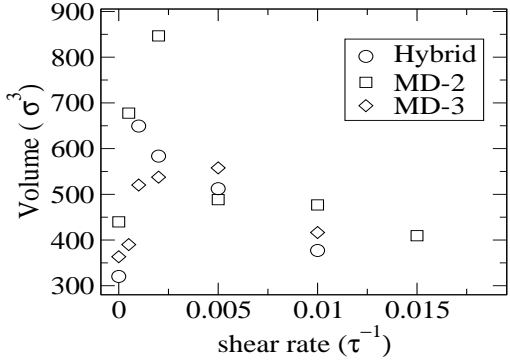}
\caption{The end-to-end volume
of the polymer as a  function of the shear rate.
The end-to-end volume is defined by the product of the
components of the end-to-end vector,
$(R^2_x\times R^2_y\times R^2_z)^{1/2}$.}
\label{volume}
\end{figure}

Figure \ref{volume} shows the end-to-end volume of the polymer,
measured as the product of the three components of the end-to-end
vector. This quantity gives an estimate of the space that the polymer
explores during its motion. 
 This volume increases for increasing shear rate and reaches a
maximum value around $\gam \sim 0.002\tau^{-1}$.  As the shear rate
is further increased the volume accessible to the polymer decreases
monotonically. This behavior of the end-to-end volume is quite similar to the
findings of Doyle {\em et al.} \cite{doyle} concerning the
amplitude of the fluctuation of the chain extension.  As the shear
rate was varied, they found that fluctuations reached a maximum size at
$\mathrm{Wi}\simeq 5.1$. Using the estimate $\tau_0 \sim 2000\tau$, we
find that the maximum end-to-end volume occurs at about
$\mathrm{Wi}\sim 4$. In fact, the size of the fluctuations
is determined by the magnitude of the 
volume made available by the polymer motion; 
or, in other words, larger fluctuations increase the explored volume.
We shall present a more detailed comparison with the
results of Doyle {\em et al.} \cite{doyle} in  future work.

There are computational costs to the hybrid method that are not
present for classical MD. These include simulating the continuum
regime and the calculations arising from the coupling procedure 
within the overlap region, {\em e.g.} particle insertion and deletion and
the evaluation of the particle stress tensor.  
For the flow considered here, the
solution of the continuum flow required around $0.01\%$ the time
needed for a LJ force calculation. 
In general, the computational time spent in simulating
the continuum region depends on the problem considered, but in any case
it will always be much smaller than the MD force evaluation for the
solvent. Furthermore, the calculation of the C-flow occurs once for
every $\sim 20$ LJ force calculations, which ensures extra savings
in computational time.  
As shown in  Appendix \ref{cpu.sec}, 
the coupling protocol, within the overlap region,
is very efficient: only  $0.01\%$ of the total 
computational time was spent in particle insertion and deletion
while around  $5\%$ in the evaluation of the particle stress tensor. 
The hybrid code as tested here needs less than half the
solvent particles, thus the overall savings in computational time is
considerable. 

\section{Conclusion}
\label{sec:con}
In this paper we have compared a newly developed hybrid MD/CFD code to
a traditional MD simulation for a single polymer tethered to  a wall
undergoing shear flow in Couette geometry. We find that the two methods give comparable
results for the conformation of the polymer within measured
uncertainty.
                                                                                
Our results indicate that the coupling protocol of the hybrid code
requires around $5\%$ of the computer time compared to the
Lennard-Jones part of the code. Most of the CPU time devoted in the ``coupling'' protocol
is spent in the evaluation of the particle momentum flux; while insertion
and extraction of particles is rather fast, taking less than $1\%$ of the overall CPU time.

This implies that, compared with a traditional MD simulation, the
amount of computational time saved by the hybrid scheme is
proportional to the volume of the simulation that is described by the
coarse-grained model (CFD). In traditional MD simulations of
interfacial phenomena finite size effects significantly alter the
local interfacial dynamics, and they can only be reduced by increasing
the volume of the simulation box that surrounds the interfacial region
of interest.  This means that most of the computational cost is likely
to be spent in the resolution of the bulk flow. In this paper we have
shown that this drawback disappears when using a proper hybrid MD-CFD
scheme. To that end, we considered a problem which is very sensitive
to small changes in the surrounding fluid environment: the motion of a
single tethered polymer under shear flow.  The excellent agreement
found in the comparisons with the full MD results indicates that the
hybrid scheme indeed eliminates finite size effects even in relatively
small systems. This means that hybrid simulations can be expected to
significantly reduce the computational cost of appropriate interfacial
problems.
                                                                                
Apart from the savings in CPU time, the hybrid scheme enables to us
gather information from all relevant time and length scales, so it
is well suited to treat multiscale problems where bulk fluid flow
plays an important r\^ole; other examples include crystal growth from
fluid phases, wetting phenomena and membrane dynamics under flow.
                                                                                
We regard the results of the present work
as an encouraging sign for future simulations, and we plan to
explore various selected interfacial systems in forthcoming research.
                                                                                
\section{Acknowledgments}
We acknowledge fruitful discussions on the hybrid formalism with
Eirik Flekk{\o}y. This research was supported by the European Commission
through a Marie Curie Fellowship (HPMF-CT-2001-01210) and by the EPSRC
RealityGrid project grant GR/R67699.  R. D-B also acknowledges support from
project BFM2001-0290.

\appendix
\section{Mass, longitudinal momentum and energy fluctuations}
\label{sec:appx}

In this work the mean solvent  flow carries no
longitudinal momentum along the $z$-direction and has a 
constant mean density. 
However, we observe that the polymer motion induces density and longitudinal
velocity fluctuations within the particle region that induce currents
of mass and longitudinal momentum travelling along the simulation box.
These perturbative currents have to be controlled at the
C$\rightarrow$P interface. We need to ensure that 
the mean mass flux across the $z=l_{CP}$ interface is zero, but in such a 
way  that any
pressure waves leave the simulation box once they reach
the C$\rightarrow$P interface. In other
words, we need to prevent any
pressure waves from bouncing back at the C$\rightarrow$P interface in the MD region.

The average number of particles crossing the C$\rightarrow$P interface
per unit time is given by $\dot{N}_{CP}=A \langle \rho
v_z\rangle_{CP}$. Zero mass flux is ensured by equating this rate
$\dot N_{PC}$ to the rate of insertion of molecules into the particle
system \cite{BusF,usher}. In the calculations presented
here we used another control equation which provides a finer control
on the particle density near the C$\rightarrow$P interface. This
approach is based on relaxing the local density at the C$\rightarrow$P
buffer to a prespecified value $\rho_O$,
\begin{equation}
\label{mfl}
\dot{N}_{CP}=\frac{V_{PC}}{\tau_{m}} \left(\langle \rho \rangle_{CP} -\rho_{O} \right)
\end{equation}
where $V_{CP}$ is the volume of the C$\rightarrow$P cell, $\langle
\rho \rangle_{CP}$ is its local the particle density averaged over
$\Delta t_{av}$ and $\tau_m$ is a relaxation time which controls the
rate at which the density fluctuations within C$\rightarrow$P cell are
smoothed out.  We set the value of $\tau_m$ slightly smaller than the
time needed by a sound wave to cross the C$\rightarrow$P cell ($\sim
O(1)\tau$).  This procedure ensures that fluctuations carrying mass
and longitudinal currents are damped at the C$\rightarrow$P cell and
do not bounce back to the inner part of the MD domain.  According to
Eq. (\ref{mfl}), particles are extracted if $\dot N_{PC}<0$ and, as
explained in \cite{BusH1}, the first particles to be extracted are
those closer to the C$\rightarrow$P interface.  If $\dot N_{PC}>0$,
new particles are inserted with a velocity extracted from a Maxwellian
distribution with mean velocity $v_y=v_z=0$ and $v_x=\gam z$ and
temperature $T=1.0$.  The insertion of particles in liquids is not a
trivial task, however, and it is addressed by the {\sc usher}
algorithm for particle insertion \cite{usher}.  The value of $\rho_O$
in Eq.(\ref{mfl}) was set to a slightly smaller value, $\rho_O=0.65$, 
than the mean density $0.8$. This reason for this choice is, first, to
alleviate the computational cost of insertion (see Appendix \ref{cpu.sec})
and, second to reduce
the amplitude of the ripples of the density profile at the
C$\rightarrow$P buffer, as shown below.  For a liquid with $\rho
\simeq 0.8$ the {\sc usher} algorithm needs around 30 iterations to
insert a LJ atom at a location where the potential energy equals the
mean specific potential energy of the system \cite{usher}, where each
iteration corresponds to the evaluation of a single-particle force. If
the density is decreased to $0.65$, it only needs about 15 iterations.
Figure 6 compares the density profile resulting from using $\rho =0.65$
in Eq. (\ref{mfl}) with that arising from a pure MD simulation. The
``hybrid'' density profile presents some ripples whose amplitude is
damped after around $3\sigma$, whereas inside the P$\rightarrow$C cell
the hybrid density profile perfectly matches the density within the
bulk.

As long as the fluid is isothermal and there are no mean pressure
gradients, the mean energy flux across the C$\rightarrow$P interface
is zero. We therefore only need to guarantee that the specific energy
of the newly inserted particles matches that of the ensemble. The
kinetic energy is matched by inserting new particles with a Maxwellian
distribution, as stated above.  In order to match the potential
energy, new particles are inserted at sites where the inter-particle
potential energy equals the chemical potential of the system, thereby
ensuring the Widom insertion criterion.

\section{Computational cost of the hybrid method}
\label{cpu.sec}

We compare the computational cost of the coupling subroutines with
those pertaining to the MD part of the hybrid scheme. This
comparison was made using the {\tt gprof}
command available in the package of the {\tt f77}
compiler. One of the parts of the hybrid scheme for which one may expect
a certain cost in computational time is particle insertion. 
Table 1 presents some results obtained for 
different shear rates and values of the density 
$\rho_O$ in Eq. (\ref{mfl}). Typically, Eq. (\ref{mfl}) requires
around 5 insertions per  time interval $\tau$ and around 15
iterations per particle (each interation involving a
single-force evaluation). Therefore, for
a time step of $\Delta t_P=0.0075 \tau$,
the insertion of new particles
needs typically about one extra force evaluation per time step. 
This number is very small when compared
with the number of force evaluations needed in the 
MD system, which is on the order of the
number of particles $N_p\sim 10^4$. 
This estimate is consistent with our findings concerning the computational
cost. As shown in table 1,  
in hybrid calculations using $\rho_O=0.65$, the time spent in the 
insertion/extraction subroutines was about $1.5 \times 10^{-4}$ times the
time spent in the force evaluation and around $0.9\times10^{-5}$
if one includes the Verlet list evaluation.
This performance confirms the extremely high
efficiency of the {\sc usher} algorithm for particle insertion. 

As a matter of fact, the dominant cost of
the hybrid scheme resides in the evaluation of the particle momentum flux.
Its cost in CPU time was about 0.06 times the cost of the force plus Verlet
list subroutines. We note that the implementation of this part of our
code was not constructed in an efficient way because we evaluated the
particle momentum flux at each MD time step. Considering that for the evaluation
of $\langle j_p\rangle$ we used measurements of $j_p$ separated by its
decorrelation time, about $0.06\tau$ \cite{BusF},
we could have measured the particle flux roughly every $10$ time steps and
further reduced that ratio by a factor 10. Finally, the time needed to solve the
diffusion equation in the continuum domain was very small compared
with the MD force subroutine, by a factor of less than $10^{-4}$. 
In general, the computational time required 
to solve the continuum system will, of course, 
depend on the specific problem solved.

\begin{table}
  \begin{tabular}{|c|c|c|c|c|c|}
\hline  
$\gam\,(\tau^{-1})$ & $\rho_O\;(\sigma^{-3})$ &  $\dot
    N_{in}\,(\tau^{-1})$ & $n_{iter}$ & $E_e$ & $\frac{CPU[insert]}{CPU[force]}$\\
\hline
  0.001 &  0.8 & 3.68 & 25.4  & 0.015 &  $2.7\,10^{-4}$\\
  0.010 &  0.8 & 3.64 & 25.9  & 0.015 & \\
  0.010 &  0.65  & 8.25& 14.7 & 0.006 & $0.9\,10^{-4}$\\
  0.005 &  0.65  & 4.34& 16.3 & 0.010 &  \\

\hline  
\end{tabular}
\caption{Details of the particle insertion in several hybrid
  simulations done at shear rate $\gam$ . Using $\rho_O$ in
  Eq. (\ref{mfl}), the average rate of particle insertion 
was $\dot N_{in}$ and the average number of iterations needed by the {\sc usher}
  algorithm to insert a new particle was $n_{iter}$. $E_e$ is the
relative error in the energy upon insertion (the 
relative difference between the target potential energy 
and the potential energy at the insertion site). In the last column we
show the ratio between the CPU time 
used by the  insertion/extraction subroutines and the CPU time
used by the force subroutine plus
the Verlet neighbor list.}
\end{table}

\bibliography{text19}
\end{document}